\documentclass[aps,twocolumn,pra,a4paper,showpacs,10pt,floatfix]{revtex4-1}
\usepackage{graphicx}
\usepackage{keyval2e}
\usepackage[caption=false]{subfig}
\usepackage{hyperref}
\usepackage{amsmath}
\usepackage{bbold}
\usepackage{color}
\allowdisplaybreaks[4]
\usepackage[ruled]{algorithm2e}
\newcommand{\ket}[1]{| #1 \rangle}

\usepackage{cleveref}
\crefname{figure}{Fig.}{Figs.}
\crefname{section}{Sec.}{Secs.}

\begin{document}

\title{Bell inequality violation in the presence of vacancies and
  incomplete measurements} \author{Kaila C. S. Hall}
\email{kaila.hall@strath.ac.uk} \affiliation{SUPA Department of
  Physics, University of Strathclyde, Glasgow, G4 0NG, United Kingdom}

\author{Daniel K. L. Oi} \affiliation{SUPA Department of Physics,
  University of Strathclyde, Glasgow, G4 0NG, United Kingdom}

\begin{abstract}
  The characterization of a quantum system can be complicated by
  non-ideal measurement processes. In many systems, the underlying
  physical measurement is only sensitive to a single fixed state,
  complementary outcomes are inferred by non-detection leaving them
  vulnerable to out-of-Hilbert space errors such as particle loss. It
  is still possible to directly verify the violation of a Bell
  inequality, hence witness entanglement of a bipartite state, in the
  presence of large vacancy rates using such an incomplete measurement
  by optimizing the measurement settings. The scheme is robust
  against imperfect \textit{a priori} state knowledge and also
  moderate amounts of error in state determination.
\end{abstract}

\date{\today}


\maketitle

\section{Introduction}

Non-ideal states and measurements are issues in experimental
implementations of quantum information processing. Many read-out
mechanisms are only sensitive to the presence of one logical state of
a qubit, the complementary result is inferred from a non-detection,
e.g. electron shelving in ion traps
\cite{BullAmPhysSoc.20.637,Science.237.612}. Such incomplete
measurements are vulnerable to vacancies or missing particles, a
potential problem in schemes such as trapped particles in optical
lattices where the initial state occupancy could be subject to
fluctuations and where detecting the presence of a particle without
disturbing its internal state is difficult~\cite{Nat.467.68}.

Under such circumstances, it is important to be able to reliably infer
important properties of the quantum system despite incomplete
measurement. Here, we show how entanglement can still be detected in
the presence of such imperfections by optimizing the observables
measured. Even for very low occupation probability, by maximizing the
classical correlation we are still able to violate the local realistic
bound by exploiting the small entangled components of the mixture. The
optimization is also tolerant of a small degree of measurement error,
we analyze the situation corresponding to imperfect detection
efficiency and dark count.

This paper is organized as follows: \cref{sec:system} describes our
system and how the state is prepared depending on the vacancy
position. We review the CHSH inequality in
\cref{sec:detectingentanglement} and apply it to the situation with
incomplete measurement and vacancies. We optimize the measurement
settings in \cref{sec:optimisingchsh} for several scenarios and
finally in \cref{sec:effectimperfections} we analyze the robustness of
the procedure to various error models.

\section{System}
\label{sec:system}

\begin{figure}[ht]
\subfloat[ ]{
\fbox{
\includegraphics[width=.45\columnwidth]{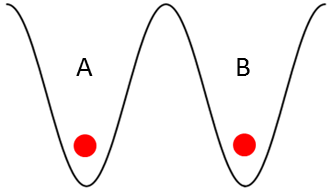}
\label{fig:boththere}
}}
\subfloat[ ]{
\includegraphics[width=.45\columnwidth]{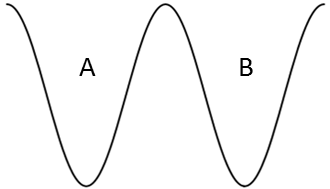}
\label{fig:nonethere}
}
\\
\subfloat[ ]{
\includegraphics[width=.45\columnwidth]{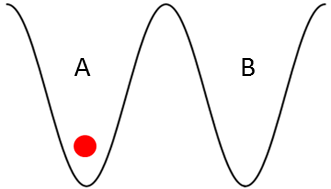}
\label{fig:leftthere}
}
\subfloat[ ]{
\includegraphics[width=.45\columnwidth]{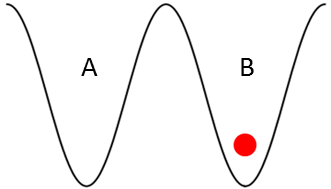}
\label{fig:rightthere}
}
\caption{(Colour online) Four possible starting
  states. \protect\subref{fig:boththere} shows the entangled case,
  $\rho_{11}$, where there are two particles in the system, one in Alice's
  site and one in Bob's this state has probability,
  $(1-p)(1-q)$. \protect\subref{fig:nonethere} shows a non-entangled
  case, $\rho_{00}$, where neither Alice nor Bob have an particle in their
  site with probability $p$. In \protect\subref{fig:leftthere} is a
  non-entangled case where only Alice has an particle, $\rho_{10}$, with
  probability $(1-p)q\frac{1-r}{2}$. In
  \protect\subref{fig:rightthere} the final non-entangled case where
  only Bob has the particle, $\rho_{01}$, with probability
  $(1-p)q\frac{1+r}{2}$.}
\label{fig:fourpossible}
\end{figure}

We consider a bipartite system where Alice and Bob may make
measurements on their respective sites as shown in
Fig.~\ref{fig:fourpossible}. Ideally, they should share a joint state
of two qubits but due to imperfections in state preparation one or
both particles may be missing \cite{Nat.467.68}. Their measurements
cannot distinguish between one of the logical states or a vacant
site. Despite this limitation, we are interested in whether we can
still detect if there is entanglement in the system.

The local state space is spanned by the states,
$\{\ket{v},\ket{0},\ket{1}\}$, representing a vacancy (no particle),
and the logical states 0 and 1 respectively. The measurement process
we consider is incomplete in the sense that it cannot distinguish
between one of the logical states of the qubit (e.g. $\ket{1}$) or its
absence, $\ket{v}$. For example, if we encode the logical states in
the energy levels of an particle
$\{\ket{0}=\ket{g},\ket{1}=\ket{e}\}$, a standard readout method is to
drive a cycling transition between $\ket{g}$ and third level $\ket{f}$
using a resonant laser~\cite{Nat.471.319}. Fluorescence is associated
with a measurement outcome of the state $\ket{0}$ whereas its absence
is mapped to the state $\ket{1}$, but this outcome is degenerate with
the absence of the particle in the first place.

\begin{figure}[ht]
\includegraphics[width=\columnwidth]{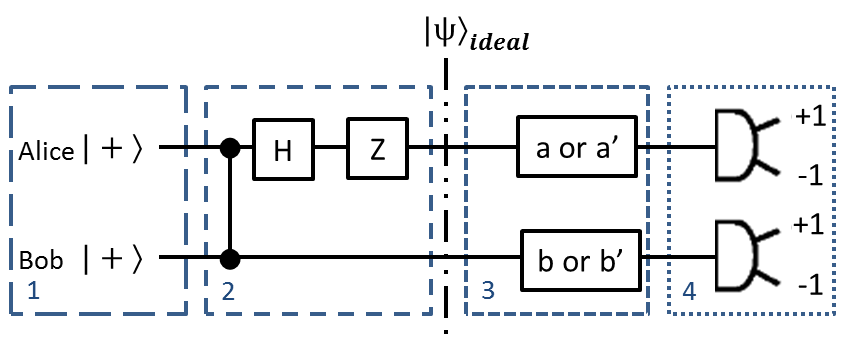}
\caption{(Colour online) Quantum circuit for experimental model.\\
  \textbf{\textcolor[RGB]{31,73,125}{1:}} Initial starting state, one of the four possible shown in \cref{fig:fourpossible}. Here we show the ideal case where $|+ \rangle=\frac{|0\rangle +|1 \rangle}{\sqrt{2}}$. \\
  \textbf{\textcolor[RGB]{31,73,125}{2:}} Operations are performed on the particles creating the fully entangled Bell state $|\psi^{-} \rangle = \frac{|01 \rangle - |10 \rangle}{\sqrt{2}}$ \\
  \textbf{\textcolor[RGB]{31,73,125}{3:}} Alice (Bob) choose to measure along axis $a(b)$ or $a'(b')$ on the Bloch sphere with some probability. These measurements are performed via an active rotation of the particle.\\
  \textbf{\textcolor[RGB]{31,73,125}{4:}} Read out of measurement
  results in fixed basis.}
\label{fig:quantumcircuit}
\end{figure}

We model the situation with an initial mixed density operator with
sectors corresponding to both particles missing $\rho_{00}$, only one
particle present $\rho_{01}$ and $\rho_{10}$, or both $\rho_{11}$. We
assume there are no coherences between sectors due the lack of a
suitable reference frame to lift the superselection rule on particle
number~\cite{PhysRev.88.101,NJPsuperselection}. Hence, we can
parameterize the initial state as
\begin{equation}
\begin{split}
\rho_{AB}(p,q,r) &= p\rho_{00} +(1-p) \rho\\ 
&=p\rho_{00} \\
+(1-p)&\left( q \left( \frac{1+r}{2}\rho_{01} 
+ \frac{1-r}{2}\rho_{10} \right) +(1-q) \rho_{11} \right)
\label{eq:rhototal}
\end{split}
\end{equation}
where $\{p,q\}$ are probabilities and $-1\le r \le +1$ characterizes
the asymmetry in the vacancy rates of Alice and Bob. In the ideal case
of no vacancy, a sequence of operations would produce the maximally
entangled singlet state
$\ket{\Psi_{ideal}}=\frac{1}{\sqrt{2}}(\ket{0,1}-\ket{1,0})_{AB}$ as
shown in Fig.~\ref{fig:quantumcircuit}. We assume that there is no
transfer population between the two sites and that the gates are
ideal. In order to detect the generation of entanglement, we may
perform various measurements. As is the case in many physical systems,
the physical measurement basis is fixed but preceding coherent
rotations allow an arbitrary choice of basis of the state.

\section{Detecting Entanglement}
\label{sec:detectingentanglement}

A sufficient condition for a state to be entangled is that it violates
a Bell inequality~\footnote{Of course, if one is able to perform
  quantum state tomography of the system, then other tests could be
  applied such as non-positivity of the partial
  transpose~\cite{PhysRevLett.77.1413},
  concurrence~\cite{PhysRevLett.80.2245,PhysRevLett.78.5022},
  etc. However, an incomplete measurement complicates the
  reconstruction process though strategies exist to accommodate
  imperfect detection by ``measuring the
  measurement''~\cite{Feito2009}. However, a violation of a Bell
  inequality does not rely on any assumptions about the form of the
  measurements or the processes hence is a more direct method of
  verification~\cite{Phys.1.195}.}~\cite{QInfo.barnett}. In the
simplest CHSH form this requires that Alice and Bob independently
choose one of two alternative measurements $\{a,a'\}$ and $\{b,b'\}$
respectively at random and assign $\pm 1$ to the two possible
outcomes~\cite{PhysRevLett.23.880,nielsenchuang}. Under the assumption
of local realism the magnitude of the correlation function $|S|
=|\langle ab \rangle + \langle ab' \rangle + \langle a'b \rangle -
\langle a'b' \rangle| \le 2$~\cite{PhysRevLett.23.880}. Quantum
mechanics allows $|S|=2\sqrt{2}$ for a maximally entangled state of
two qubits~\cite{LiMP.4.93}. In our case, a vacancy adds a third
possible value but due to the measurement process, it is
indistinguishable from the logical state $\ket{1}$. We consider our
measurement as an incomplete projection with one outcome $+1 \iff
|0\rangle$ and the other degenerate, $-1 \iff |1\rangle \text{ or } |v
\rangle$. We can represent the measurement settings $a,a',b,b'$ as if
we had a qubit but calculate the outcome probabilities to include the
effect of vacancies.  Hence we would like to find measurement settings
that lead to a violation of local realism for a large range of vacancy
probability.

\subsection{Bound on the CHSH value}
\label{sec:analysischsh}

We first note that $S$ is bounded as
\begin{equation}
|S| \leq \alpha 2 + (1-\alpha) 2\sqrt{2},
\label{eq:upperbound}
\end{equation}
where $0 \leq \alpha \leq 1$ is the probability of having a separable
state. The entangled component results in a maximum of $2 \sqrt{2}$
due to the Tsirelson's bound~\cite{LiMP.4.93} and the separable state
gives a maximum of $2$ for classical correlation. For any $\alpha \ne
0$ there exists the possibility of a violation of the CHSH inequality
and we will see how close we can approach this bound with incomplete
measurements.

\subsection{Simplification of the CHSH Expression}

The measurement settings $a,a',b,b'$ in the ideal case of
$\{p=0,q=0,r=0\}$ can be written in terms of projectors and their
complement on a qubit, $\cos{\frac{\theta}{2}} \ket{0} +
e^{i\phi}\sin{\frac{\theta}{2}}\ket{1}$ defining a direction on the
Bloch sphere where we have denoted
$\ket{0}=\ket{+Z}$~\cite{nielsenchuang}. As we model the physical
measurement as a projection along a fixed axis, these settings are
implemented by preceding unitary rotations that act upon the
$\ket{0},\ket{1}$ components but do not affect $\ket{v}$. Each setting
is thus represented by $\theta$ and $\phi$, e.g. $\theta_{a},\phi_{a}$
for measurement setting $a$. This allows an analytical express for
$S$:
\begin{equation}
S=2p+(1-p)S',
\end{equation}
where $p$ is the probability of $\rho_{00}$ ($\ket{v,v}_{AB}$)
occurring, and $S'$ is where there is at least one particle,
\begin{equation}
\begin{split}
S'=& q \Big((r-1)\cos{\theta_{a}}- (r+1)\cos{\phi_{b}}\sin{\theta_{b}} \Big) \\
+\frac{(q-1)}{2}&\Big[ \cos(\theta_{a}-\theta_{b})(1+\cos(\phi_{a}-\phi_{b}))\\
&+\cos(\theta_{a}+\theta_{b})(1-\cos(\phi_{a}-\phi_{b}))\\
&+\cos(\theta_{a}-\theta_{b'})(1+\cos(\phi_{a}-\phi_{b'}))\\
&+\cos(\theta_{a}+\theta_{b'})(1-\cos(\phi_{a}-\phi_{b'}))\\
&+\cos(\theta_{a'}-\theta_{b})(1+\cos(\phi_{a'}-\phi_{b}))\\
&+\cos(\theta_{a'}+\theta_{b})(1-\cos(\phi_{a'}-\phi_{b}))\\
&-\cos(\theta_{a'}-\theta_{b'})(1+\cos(\phi_{a'}-\phi_{b'}))\\
&-\cos(\theta_{a'}+\theta_{b'})(1-\cos(\phi_{a'}-\phi_{b'})) \Big].
\end{split}
\label{eq:Snop}
\end{equation}
Since $\ket{v}$ always gives the result $-1$, the $\rho_{00}$
component results in the maximum classical correlation independent of
the angles $\theta$ and $\phi$. As long as $p<1$ and $|S'|>2$, then a
CHSH violation is possible. Hence, for the purposes of optimization,
we need only consider $S'$.

\section{Optimizing measurement settings}
\label{sec:optimisingchsh}

For a given state with parameters $p,q$ and $r$ we seek measurement
setting that maximize $S'$ (\eqref{eq:Snop}), hence $S$. Without loss
of generality, it is simple to show that it is sufficient to set
$(\phi_{j}-\phi_{k} )=0,\pi \mod 2\pi$ to achieve this, i.e. perform
measurements only in the $X-Z$ plane. We choose $\phi_{a}=
\phi_{a'}=\phi_{b}=\phi_{b'}=0$ leading to the simplified expression
\begin{equation}
\begin{split}
S'= q \Big(&(r-1)\cos{\theta_{a}}-(r+1)\sin{\theta_{b}} \Big) \\
+(q-1)&\Big[ \cos(\theta_{a}-\theta_{b})+\cos(\theta_{a}-\theta_{b'})\\
&+\cos(\theta_{a'}-\theta_{b})-\cos(\theta_{a'}-\theta_{b'}) \Big].
\end{split}
\label{eq:Snop2}
\end{equation}
When Alice and Bob share a spherically
symmetric singlet state $(q=0)$, $S'$ depends only on the relative
differences between the $\theta$'s, hence it is invariant under
bi-local rotations ($U_A\otimes U_B$, $U_A=U_B$) as expected. However,
for $(q>0)$ the first term in Eq.~\eqref{eq:Snop} results in a
preferred direction of measurement settings to achieve the maximum
value.

\subsection{Equal and independent vacancy rate}
\label{sec:eqindepopt}

\begin{figure}[ht]
\includegraphics[width=\columnwidth]{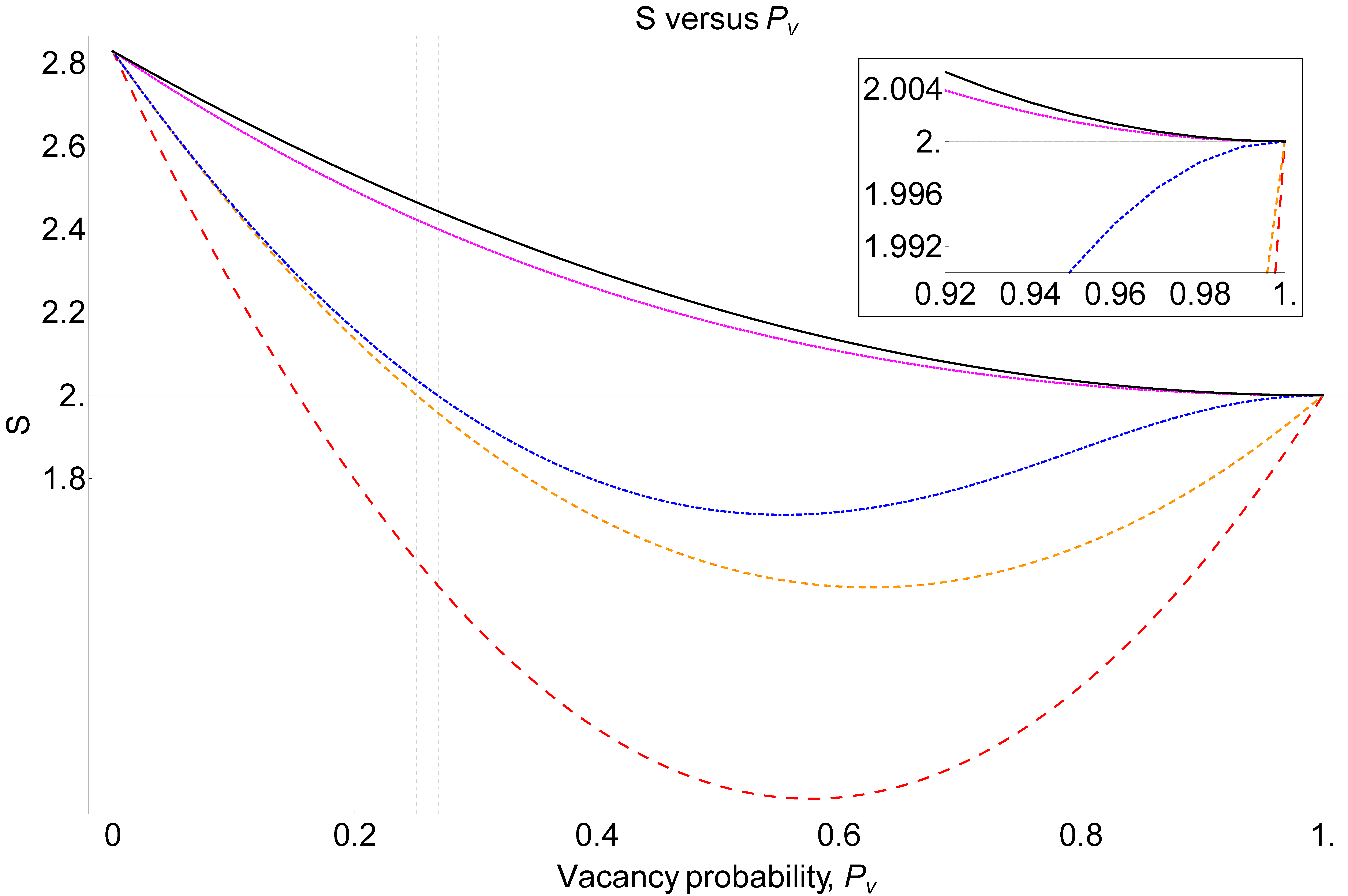}
\caption{(Colour online) Comparison of CHSH values for various
  optimizations. The black solid line is the upper bound
  given by Eq.\eqref{eq:Snop}\\
  \textbf{No optimization} (Red dashed line) Vacancy probability
  above which entanglement not detected $P_{v_0}^{crit}\approx 0.153$.\\
  \textbf{Step 1} (Orange dashed line) $P_{v_1}^{crit} \approx 0.251$.
  The optimal angle of rotation, $\theta$ is $\frac{5\pi}{8}$ for
  all $P_{v}$.\\
  \textbf{Step 2} (Blue dot--dashed line) $P_{v_2}^{crit} \approx
  0.269$.  The optimal angles of rotation for Alice and Bob's pairs of
  axes is
  different for each value of $P_{v}$.\\
  \textbf{Step 3} (Magenta dotted line) $P_{v_3}^{crit} <1$. Again the
  optimal angles of rotation of each of Alice and Bob's axes are
  different for each value of $P_{v}$.}
\label{fig:chshjointgraph}
\end{figure}

As an example we first consider the case where each site has an equal
and independent vacancy probability $P_{v}$ leading to $\rho_{AB}
(P_{v}^{2},\frac{2P_{v}}{(1+P_{v})},0)$. For different values of
$P_{v}$ we can numerically optimize the measurement settings to
maximize the value of $S'$ and we investigate the effect of various
stages of optimization (\cref{fig:chshjointgraph}).

\begin{description}

\item[No Optimization]\label{step0} Using the conventional measurement
  settings (\cref{fig:conventionalaxes}) we obtain $S$ as shown in
  \cref{fig:chshjointgraph} (Red dashed line). Above the critical
  value of $P_{v_0}^{crit}\approx 0.153$ we are unable to obtain a
  violation.

\item[Step 1]\label{step1} We first allow a bi-local rotation
  around the $Y$ axis by the same angle, i.e. redefining the global
  axis (\cref{fig:chshthetaaxes}), this extends the critical value to
  $P_{v_1}^{crit}\approx 0.251$ (\cref{fig:chshjointgraph}, Orange
  dashed line). The optimal rotation of the conventional measurement
  settings is $\frac{5\pi}{8}$ for all $P_{v}$ which can be understood
  from Eq.~\eqref{eq:Snop2}, the bi-local rotation maximizes the first
  term but does not affect the second. As this optimization does not
  depends on $q$ the optimal rotation is the same for all values of
  $P_{v}$.

\item[Step 2]\label{step2} We rotate the the local axes independently but
  keep the relative angles of Alice's settings ($a,a'$) and the
  relative angles of Bob's settings ($b,b'$) the same
  (\cref{fig:chshalphabetaaxes}). This leads to $P_{v_2}^{crit}\approx
  0.269$ (\cref{fig:chshjointgraph}, Blue dot--dashed line), however
  the optimal rotations are now $P_{v}$ dependent.

\item[Step 3]\label{step3} Finally, we optimize each measurement
  setting individually (\cref{fig:chshallaxes})~\footnote{As
    previously noted, it is only necessary to optimize in the X-Z
    plane. Optimization over the entire Bloch sphere yields the same
    maximum of $S'$ as expected} leading to $P_{v_3}^{crit}$
  arbitrarily close to $1$ (\cref{fig:chshjointgraph}, Magenta dotted
  line). If there is some probability that there are particles in the
  system, then in principle we can detect at least a small amount of
  entanglement. Again the optimal angles of rotation are dependent
  upon $P_{v}$.

\end{description}

These optimizations progressively increase the value of $P_{v}^{crit}$
to a point where we will always detect entanglement provided
$P_v<1$. In the case where $P_{v}\to 1$, the fully optimized
measurement angles approach $\theta_{a}=3.139$, $\theta_{a'}=1.048$,
$\theta_{b}=4.715$ and $\theta_{b'}=0.523$.  Inserting these into
Eq.~\eqref{eq:Snop2} we see that these measurement settings produce a
value of very close to $2$ for the separable part of the system and
but still provides enough of a violation from the entangled component
to produce a total value above 2.

\begin{widetext}

\begin{centering}
\begin{figure}[ht]
  \subfloat[No optimization. The conventional settings used in the
  CHSH inequality test on the Bloch sphere.]{
\includegraphics[width=0.4\columnwidth]{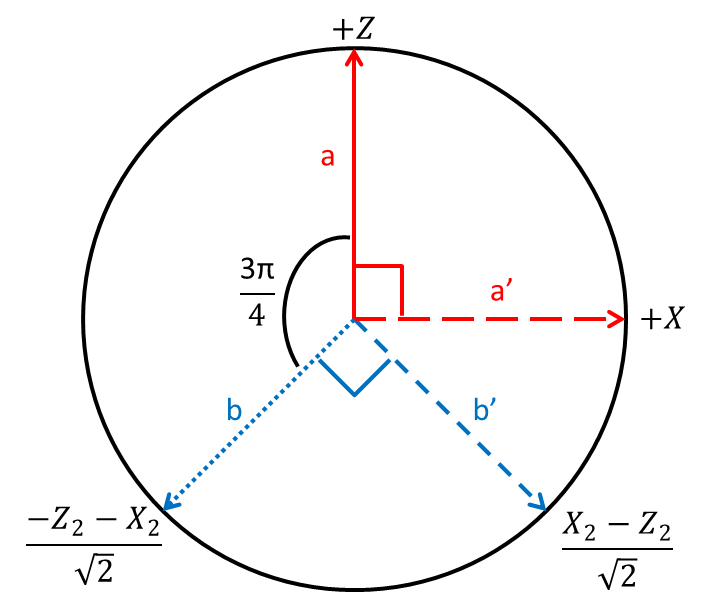}
\label{fig:conventionalaxes}
}
\qquad
\subfloat[Step 1. All settings rotated around the $Y$ axis
of the Bloch sphere by the same amount.]{
\includegraphics[width=0.4\columnwidth]{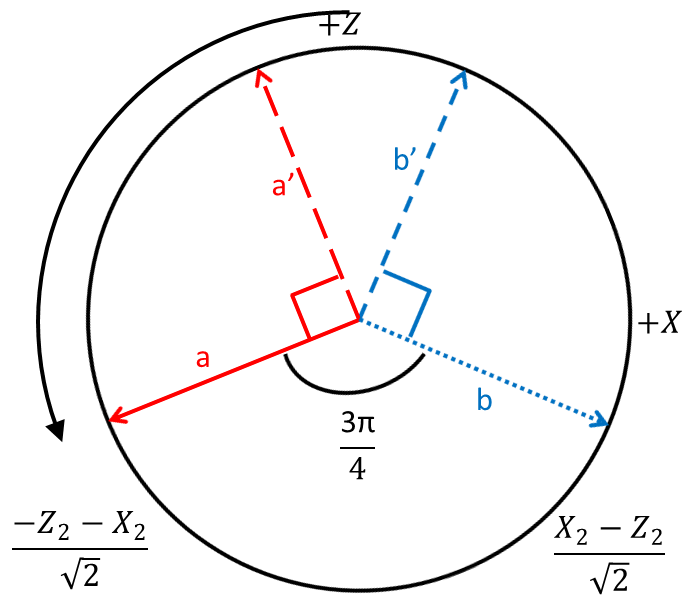}
\label{fig:chshthetaaxes}
}
\\
\subfloat[Step 2. Alice and Bob's axes are rotated independently
around the $Y$ axis of the Bloch sphere, keeping the relative angles
of each pair.]{
\includegraphics[width=0.4\columnwidth]{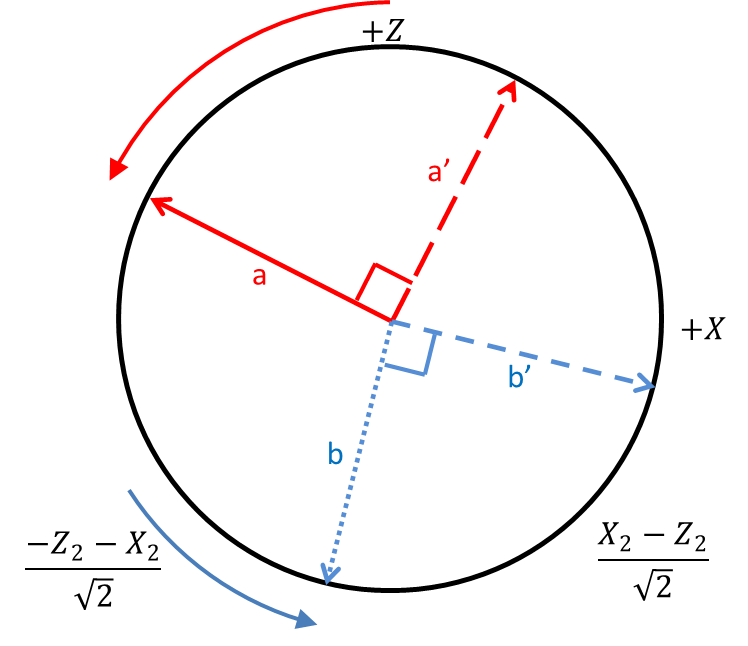}
\label{fig:chshalphabetaaxes}
} \qquad
\subfloat[Step 3. Each measurementment is allowed to vary
independently.]{
\includegraphics[width=0.4\columnwidth]{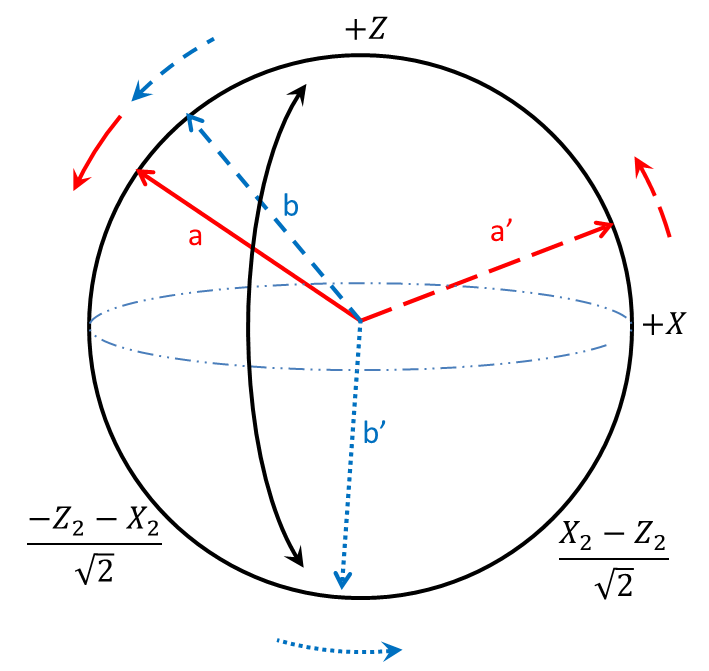}
\label{fig:chshallaxes}
}
\caption{(Colour online) Alice's choice of measurement settings $a$
  and $a'$ are shown in red, Bob's choice of measurement settings $b$
  and $b'$ in blue. These measurements are performed by an active
  rotation of the particle then a measurement in a fixed basis.}
\end{figure}
\end{centering}

\end{widetext}

\subsection{More general $\rho(p,q,r)$ states}

The optimal measurement settings are not able to saturate the bound in
Eq.~\eqref{eq:upperbound} under the equal and independent vacancy rate
model. This requires measurement settings that simultaneously give $2$
for the separable part and $2\sqrt{2}$ for the maximally entangled
component, mirroring the right hand side of Eq.~\eqref{eq:upperbound}.
We now look at more general states to see whether it can be achieved,
e.g. allowing differing and correlated vacancy rates between the two
sites.

\begin{figure}[ht]
\includegraphics[width=\columnwidth]{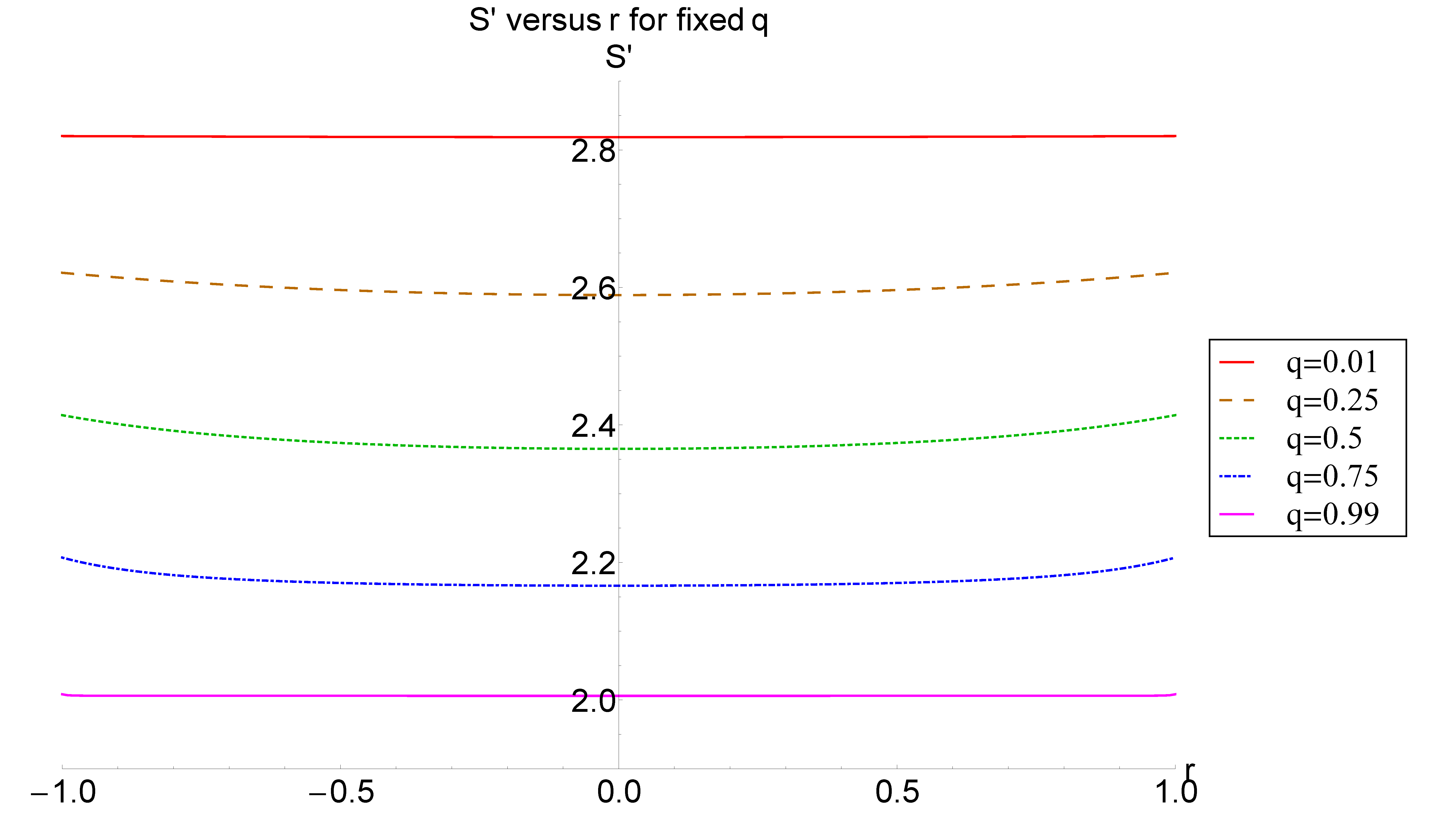}
\caption{(Colour online) $S'$ versus $r$ with a fixed $q$ value. The
  optimal measurement settings for $r=-1$ are the same for all $q$,
  this is also true of $r=+1$ .}
\label{fig:changingr}
\end{figure}

Varying $r$ (\cref{fig:changingr}), which controls the symmetry in the
one particle sector, we find that it is better to have a completely
asymmetric system, $r=\pm 1$, to produce a larger $S'$. This can be
understood from the description of the system $\rho
(0,q,1)=q\rho_{01}+(1-q)\rho_{11}$, to saturate
Eq.~\eqref{eq:upperbound} we can choose settings to produce
$2\sqrt{2}$ on the $\rho_{11}$ component up to a bi-local rotation due
to the spherically symmetric nature of the singlet state. Hence, this
rotation can be chosen so that these measurement settings acting on
$\rho_{01}$ gives $2$, the maximum correlation for a non-entangled
state. In this way states with $r=\pm1$ can saturate the bound in
Eq.~\eqref{eq:upperbound}.

The dip we see as $r \to 0$ in \cref{fig:changingr} is due to the
angles for the entangled and non-entangled parts no longer matching so
a compromise is made between them leading to a reduced $S'$.

\section{Effects of additional imperfections}
\label{sec:effectimperfections}

We have found that in principle we can always detect entanglement if
it is present. But in experiment, additional uncertainty and noise can
arise and we look at two sources of imperfections that may affect our
results.

\subsection{Robustness to state knowledge}

To use most of the optimization steps in \cref{sec:eqindepopt} it is
necessary to have prior knowledge of $p$, $q$ and $r$. In the absence
of precise knowledge of these parameters, this will reduce our ability
to detect a violation of the CHSH inequality. As an example, we
consider the case where our estimate of $q$ may be inaccurate. For
simplicity, we will assume that our estimate of $r$ is correct and we
set $r=0$ as this produces the smallest optimum value of $S'$.

\begin{figure}[ht]
\centering
\includegraphics[width=\columnwidth]{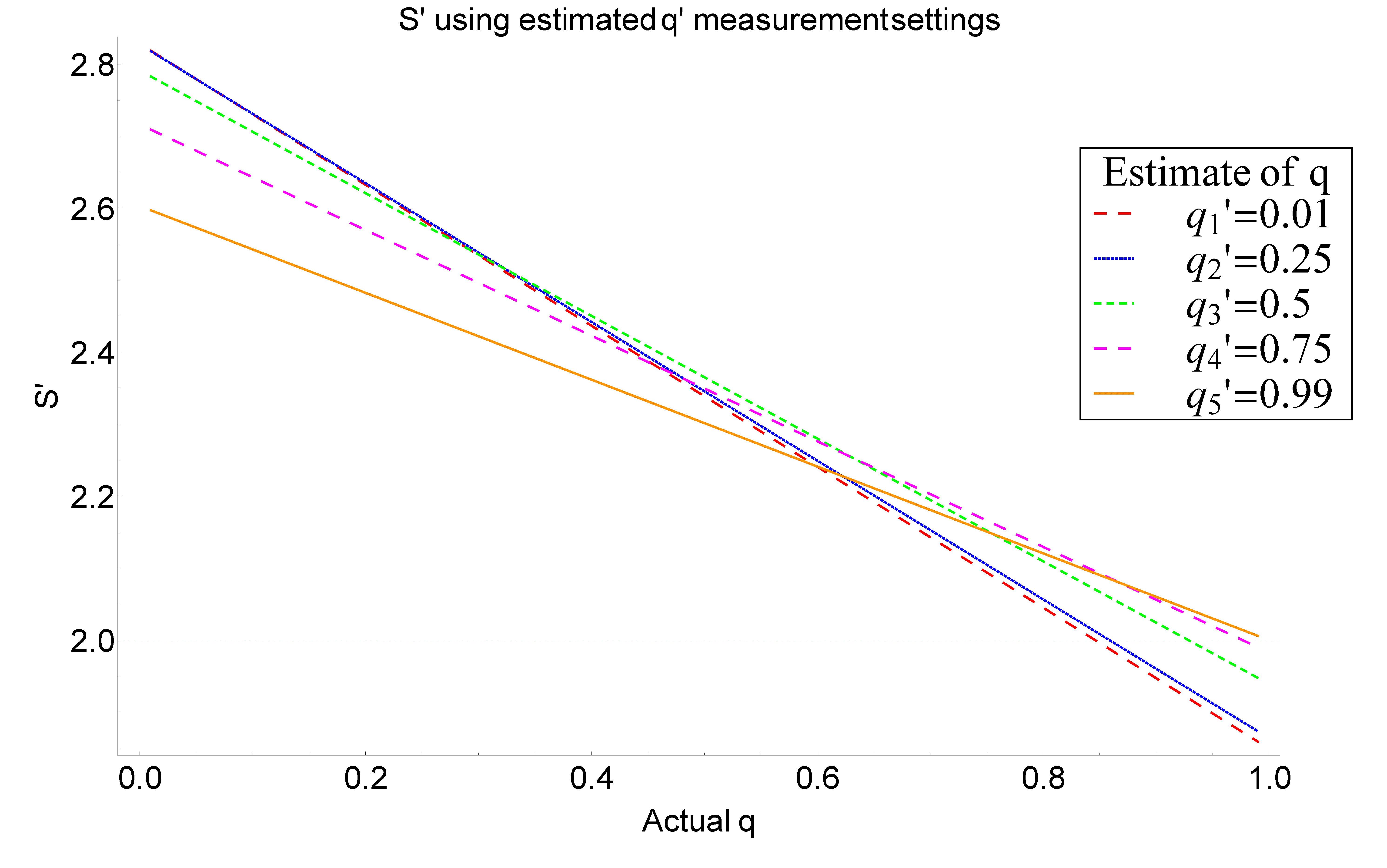}
\caption{(Colour online) $S'$ for estimated $q'$ fixed measurement
  settings. Using the optimal measurement settings calculated for
  $q'_{1} \dots q'_{5}$ with $r=0$ we optimize $S'$ over the range of
  $q$ individually for these five fixed measurement settings. Choosing
  the worst case scenario where $q'=0.99$ (Orange solid line) shows
  entanglement over the whole range of $q$.}
\label{fig:graphcomparingassumedq}
\end{figure}

We denote our uncertain estimate of the true value of $q$ as
$q'$. Should the true value of $q$ be different from $q'$ then our
measurement settings will not be optimum for $q$. The question becomes
whether we can still detect entanglement for these non-optimum
settings. \cref{fig:graphcomparingassumedq} shows the effect of
assuming different values of $q'$ on $S'$. We see that it is better to
be pessimistic on our estimate of $q'$ i.e. if we choose optimum
values for a higher value of $q'$ we can detect a violation of the
CHSH inequality for a larger range of $q's$.

\subsection{One-sided error}
\label{sec:onesidederror}

So far we have assumed that the detection process is $100\%$ efficient
even though it is incomplete. However real detectors can suffer from
dark count and inefficiency leading to error in identifying the
correct result.\cite{NatPhoto.3.696}.

\subsubsection{Photons/ detector inefficiency}

In the case of resonance fluorescence the measurement relies on
efficient capture of all the scattered
photons~\cite{BullAmPhysSoc.20.637,Science.237.612,Wineland:80}. Insufficient
captured solid angle combined with detectors that may not register
every photon that falls upon them can lead to mis-identification of
the fluorescent state with the dark state or vacancy. This
induces a one-sided measurement error where
\begin{equation}
\begin{split}
P(+1| \, |0\rangle )&=1-P(-1| \, |0\rangle )=\eta \\
P(-1 | \, |1\rangle) &= P(-1| \, |v\rangle)=1
\end{split}
\end{equation}
where $0\le (1-\eta)\le 1$ is the probability of incorrectly identifying
the $\ket{0}$ state as $\ket{1}$ or $\ket{v}$, with the ideal case
$\eta=1$ under the model in \cref{sec:eqindepopt}. This error does not
change the measured classical correlation due to the $\ket{v,v}$
components, hence only $S'$ needs to be considered as before.

\begin{figure}[ht]
\includegraphics[width=\columnwidth]{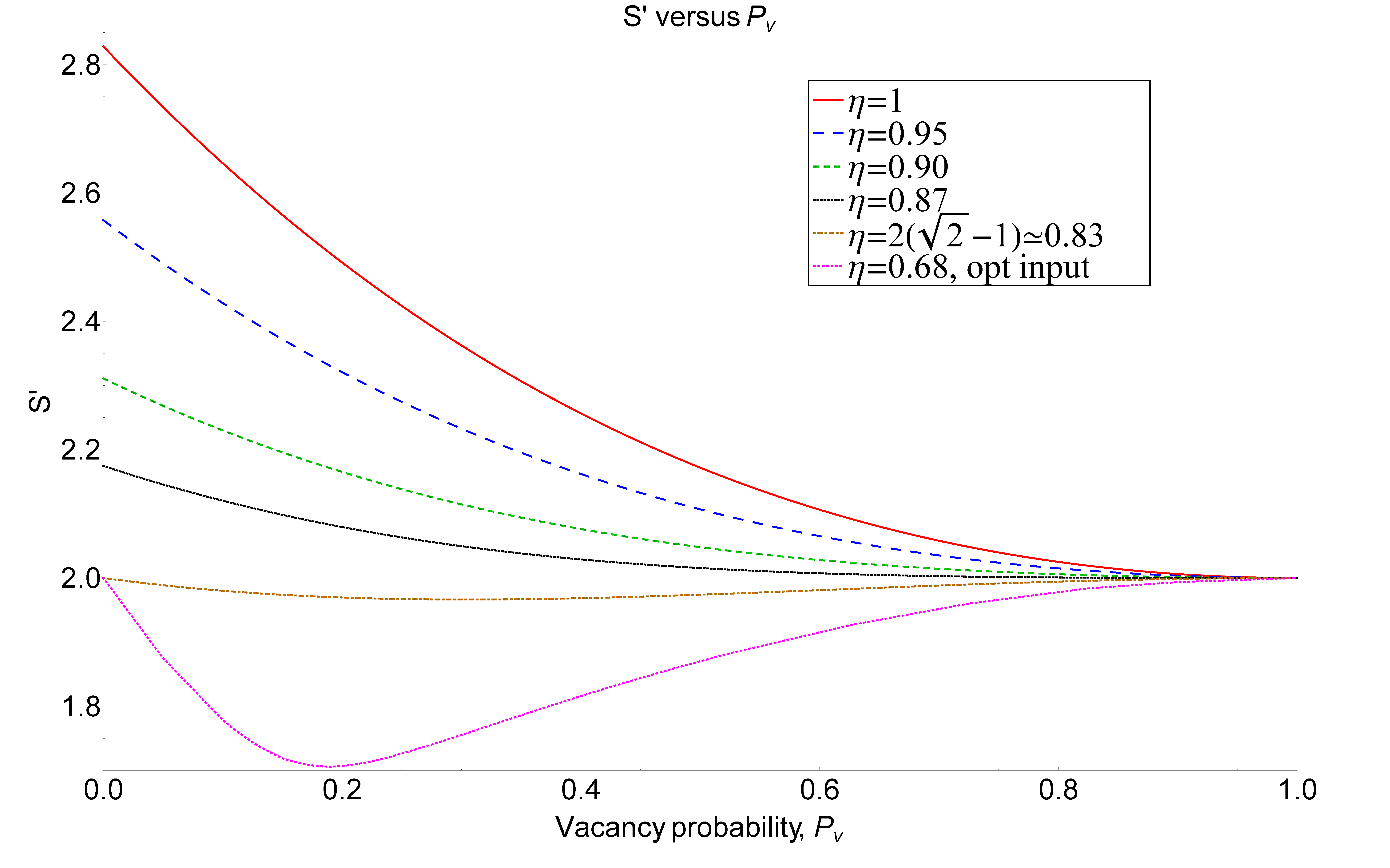}
\caption{(Colour online) $S'$ value with inefficient detection. Under
  the model in \cref{sec:eqindepopt}, $\eta=0.869$ is the largest error
  rate that will allow the value of $S$ to remain over $2$ where
  $P_{v} < 1$, and $\eta > 2(\sqrt{2}-1)$ is the limit that will allow
  a violation of the CHSH inequality for some values of $P_{v}$. By
  optimizing the initial input state as well, it is possible to
  decrease $\eta$ to as low as $0.68$ (Magenta line) for $P_v=0$.}
\label{fig:chshgamma}
\end{figure}

As $\eta$ decreases the CHSH violation reduces as expected
(\cref{fig:chshgamma}), however it is always possible to detect
entanglement across the full range of $0 \leq P_{v} < 1$ when $\eta
\gtrsim 0.869$. When $2(\sqrt{2}-1) < \eta \leq 0.869$, then the we
obtain a violation for some values of $P_{v}$. No violation is found,
even for $P_{v}=0$, when $\eta \leq 2(\sqrt{2} -1)$.

\begin{figure}[ht]
\includegraphics[width=\columnwidth]{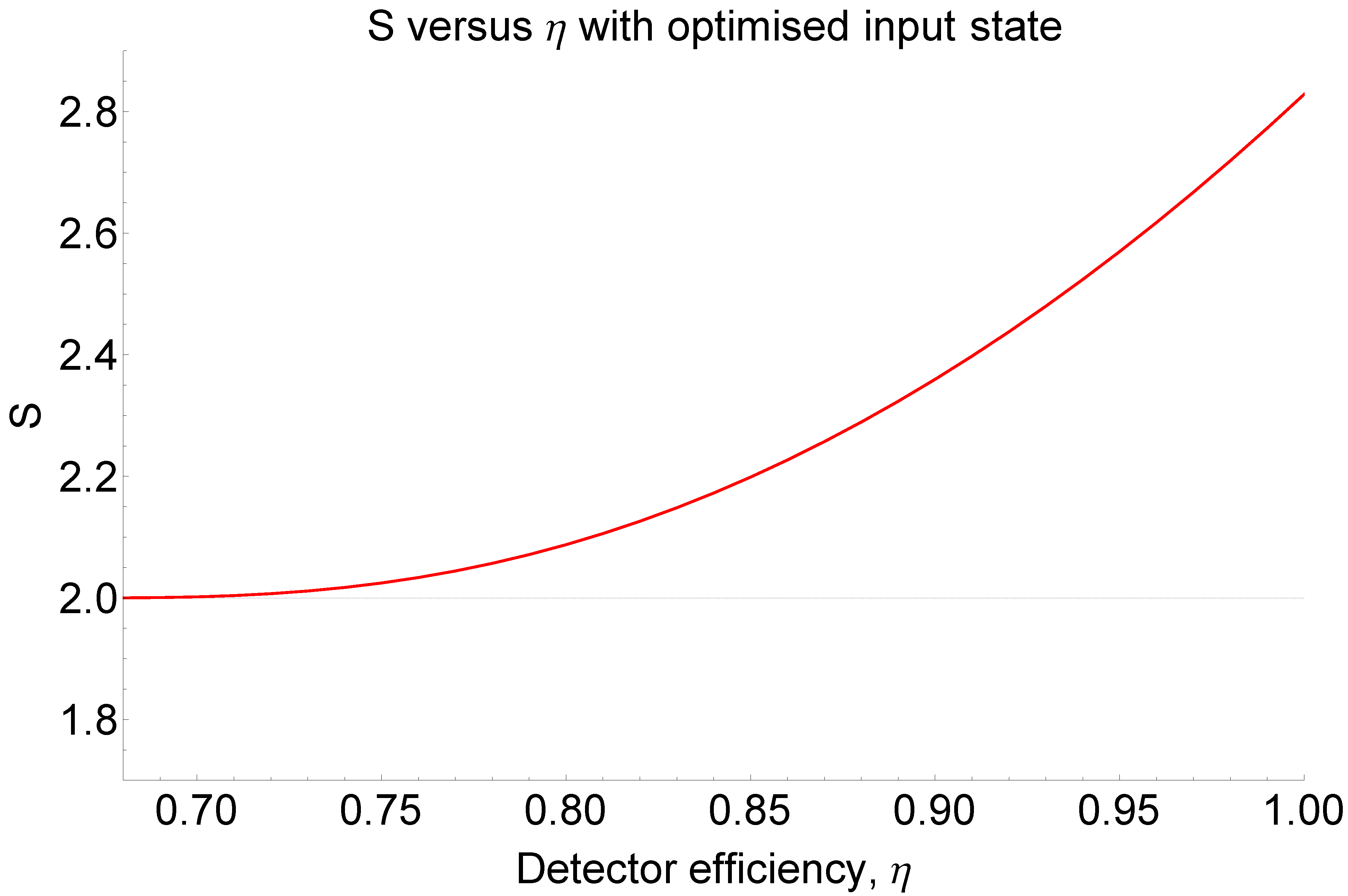}
\caption{(Colour online) $S$ versus $\eta$ with $P_{v}=0$ for
  optimized input state. For $\eta=0.68$, $S$ just exceeds $2$ hence,
  in principle a violation could be obtained.}
\label{fig:etagraph}
\end{figure}

Having a sufficiently high detection efficiency is a requirement for
closing the detection loophole~\cite{PhysRevD.2.1418,
  PhysRevA.23.3003,PhysRevD.35.3831,1751-8121-47-42-424003}. For
example, Garg and Mermin~\cite{PhysRevD.35.3831} found detector
efficiency must be higher than $2(\sqrt{2} -1)$, this value is
coincidental as the assumptions and the measurement scenarios differ
with our case. This is demonstrated by further optimization where the
initial state (\cref{fig:quantumcircuit}, Box 1) is allow to vary as
$\ket{\psi_\tau}_A\ket{\psi_\tau}_B$,
$\ket{\psi(\tau)}=\cos(\tau)\ket{0}+\sin(\tau)\ket{1}$. For example,
we have found states and settings that show a CHSH violation for
$\eta=0.68$ (Magenta line in \cref{fig:chshgamma}, and
\cref{fig:etagraph} for no vacancies).

The measurement settings with these input states no longer sit in the
$X-Z$ plane but we can specify them as a unitary rotation of the
$\ket{+Z}$ direction by a rotation $[\alpha,\theta,\phi]$ where
$\alpha$ is a rotation angle about a specified axis $(\theta,\phi)$ in
polar coordinates on the Bloch sphere. For $P_{v}=0$ and $\eta=0.68$
the optimum settings are $\tau \approx 6.11$, $a \approx
[4.16,1.95,4.14]$, $a'\approx [2.21,1.38,0.77]$, $b\approx
[2.83,1.56,1.58]$ and $b' \approx [2.46,1.48,0.72]$. We have not yet
explored further optimization of the input state.

\subsubsection{Error due to dark count}

\begin{figure}[ht]
\includegraphics[width=\columnwidth]{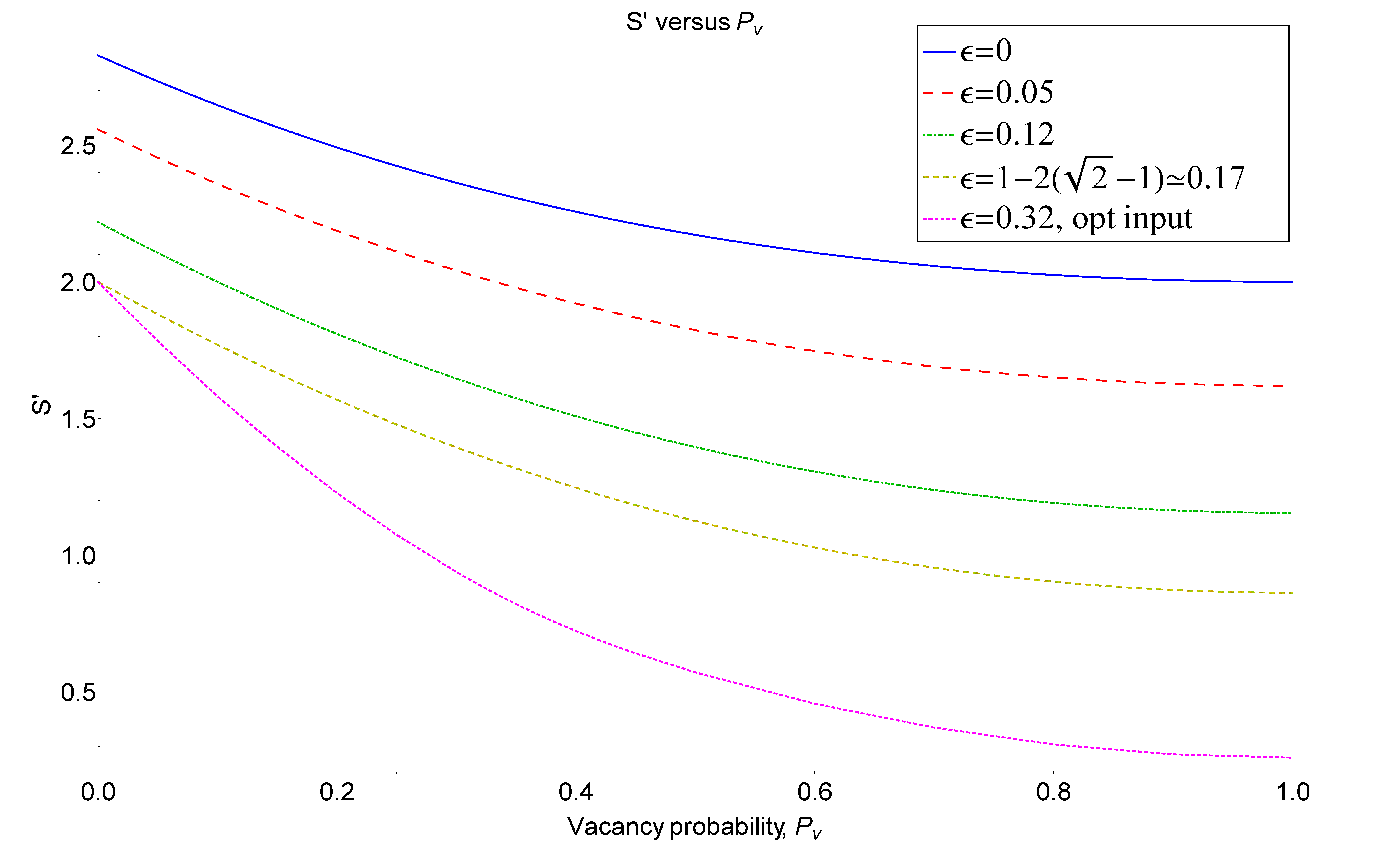}
\caption{(Colour online) Effect of dark count induced error on $S$.
  When $\epsilon > 0.17$ a CHSH violation is not for any value of
  $P_{v}$ for a $|++\rangle_{AB}$ input state. With an optimized input
  state the threshold for violation becomes $\epsilon \leq 0.32$, a
  significant improvement.}
\label{fig:chshdarkcount}
\end{figure}

Dark counts can lead to the opposite one-sided error (assuming
$\eta=1$) where the dark state $\ket{-1}$ or $\ket{v}$ can be
misidentified as $\ket{0}$,
\begin{equation}
\begin{split}
P(+1| \, |0\rangle )&=1-P(-1| \, |0\rangle )=1 \\
P(-1 | \, |1\rangle) &= P(-1| \, |v\rangle)=1-\epsilon
\end{split}
\end{equation}
where $0 \leq \epsilon \leq 1$ is the error due to dark counts, with
the ideal case $\epsilon=0$ under the model in
\cref{sec:eqindepopt}~\footnote{The use of cooled
  detectors~\cite{1487059,307321,OPL:8171925,OPL:8057345,Widenhorn2002}
  should be able to reduce dark counts to negligible levels and so
  eliminate the effect of this error channel except for extremely high
  vacancy rates.}. From \cref{fig:chshdarkcount} we can see that if
$\epsilon\geq 1-2(\sqrt{2}-1)$ it is no longer possible to detect a
violation of the CHSH inequality for any value of $P_{v}$. And for any
$\epsilon > 0$ it is no longer possible to detect a violation of the
CHSH inequality across the whole range of $P_{v}$. It is clear that
the errors induced by dark count is much more destructive to the $S'$
value.

Again, by varying the input state we can increase $\epsilon \leq
0.32$ (\cref{fig:chshdarkcount} Magenta line) and still obtain a
violation for $P_v=0$. Once again the optimized measurement settings
do not sit in the $X-Z$ plane of the Bloch sphere and for $P_{v}=0$
and $\epsilon=0.32$ the optimum settings are $\tau \approx
3.31$, $a \approx [3.82,4.18,3.73]$, $a' \approx [4.18,1.62,1.95]$, $b
\approx [5.95,4.39,1.56]$ and $b' \approx [1.45,3.68,4.70]$.

\section{Conclusion}

In this paper we have explored the effects that incomplete measurement
and vacancies have on the detection of entanglement using the CHSH
inequality in a simple two site system. For a model with equal and
independent vacancy rates, we showed that in principle it is possible
to always detect a CHSH violation if entanglement is present, hence
witness entanglement, by the use of optimized measurement
settings. Conventional CHSH measurement settings are only able to show
a violation for low vacancy rates.

As a function of the entanglement fraction of the state, we are able
to saturate the bound of the CHSH value for mixtures by considering
asymmetric states, ones where one side does not have any
vacancies. This allows the measurement settings to simultaneously
maximize the CHSH values for both the separable and maximally
entangled fraction, something not possible for symmetric vacancy
rates. This asymmetric system may be difficult to produce in practice
or may not naturally occur, but an interesting question arises from
this result, would having access to a complete local measurement that
can distinguish the presence of a particle allow us to saturate the
bound in Eq.~\eqref{eq:upperbound} for more general states?

Our procedure using modified CHSH measurement settings is robust when
either state knowledge or the measurement process itself are prone to
error. In the case of uncertainty of the vacancy rate, a conservative
estimate and associated choice of optimum settings will result in a
(reduced) CHSH violation over a large range of actual vacancy
rates. Choosing such optimum settings for a large vacancy rate will
still allow witnessing entanglement over most of the vacancy rate
range despite uncertainty as to the actual fraction of the entangled
state in the mixture.

When the measurement process is prone to a one-sided error, a CHSH
violation can still be obtained for moderate levels of inaccuracy. For
zero vacancy rates, the limit on the error levels for which violation
can be obtained is the same for both types of one-sided
error. However, in the presence of vacancies, errors due to dark count
are much more detrimental to the value of the violation and
considerably reduces the class of states for which entanglement can be
witnessed. By optimizing the input state, the tolerable error is
considerably increased. More robust states and measure settings are
obtained by using initial states that do not result in a maximally
entangled state in the ideal case. This trades the maximal violation
on the entangled fraction for greater classical correlation on the
separable part that are more resistant to the one-sided error
considered here.

The precise values of detector inefficiency differ from those obtained
for closing the detector loophole due to the experimental scenarios
considered. In our situation, we assign a result to all experimental
runs and do not discard undetected events as these are assigned the
same value as the dark state. The emphasis of this work is also
different, we are interested in witnessing entanglement and not
necessarily closing loopholes in the violation of local-realism.

Other types of error, including those in state preparation and the
implementation of the measurements rotations, will of course reduce
further the class of states for which a CHSH violation (hence
entanglement) can be directly witnessed. If the experimental setup has
already been characterized, CHSH tests could be performed on the
corrected or inferred data in order to detect entanglement though this
adds a layer of complication. An analysis of more general error models
will be left for future work.

\bibliographystyle{apsrev4-1}


%

\end{document}